\documentclass[aps,eqsecnum,superscriptaddress,12pt, floatfix]{revtex4}
\usepackage[T1]{fontenc}
\usepackage[latin1]{inputenc}
\usepackage{graphicx}
\usepackage{axodraw}
\usepackage{amssymb}
\usepackage{epsfig}

\def\lsim{\raise0.3ex\hbox{$<$\kern-0.75em\raise-1.1ex\hbox{$\sim$}}}
\def\gsim{\raise0.3ex\hbox{$>$\kern-0.75em\raise-1.1ex\hbox{$\sim$}}}

\begin{document}
\title{Testing the Scale Dependence of the Scale Factor $\sigma_{\rm eff}$ in Double Dijet Production at the LHC}
\author{Svend Domdey$^{1,2}$\footnote{Email: domdey@tphys.uni-heidelberg.de}, Hans-J\"urgen Pirner\footnote{Email: pir@tphys.uni-heidelberg.de}}
\address{Institut f\"ur Theoretische Physik, Philosophenweg 19, D-69120 Heidelberg,
Germany}
\author{Urs Achim Wiedemann\footnote{Email: Urs.Wiedemann@cern.ch}}
\address{Department of Physics, CERN, Theory Division,
CH-1211 Gen\`eve 23, Switzerland}

\begin{abstract}
The scale factor $\sigma_{\rm eff}$ is the effective cross section used to characterize 
the measured rate of inclusive double dijet production in high energy hadron
collisions. It is sensitive to the two-parton 
distributions in the hadronic projectile. In principle, the scale factor 
depends on the center of mass energy and on the minimal 
transverse energy $E_{\rm T, min}$ of the jets contributing to the 
double dijet cross section. Here, we point out that proton-proton collisions
at the LHC will provide for the first time experimental access to these scale dependences
in a logarithmically wide, nominally perturbative kinematic range
$10\, {\rm GeV} \lsim E_{\rm T, min} \lsim 100\, {\rm GeV}$. This constrains the dependence 
of two-parton distribution functions on parton momentum fractions and parton localization in impact parameter space. Novel information is to be expected about the transverse growth of hadronic distribution functions in the
range of semi-hard Bjorken $x$  ($0.001 \lsim x \lsim 0.1$) and high resolution $Q^2$. 
We discuss to what extent one can disentangle different pictures of the $x$-evolution of two-parton distributions in the transverse plane by measuring  
double-hard scattering events at the LHC.
\end{abstract}

\maketitle

\section{Introduction}
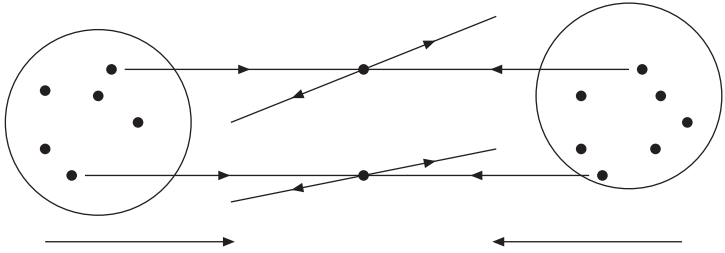
\begin{figure}[t]
\begin{center}
\begin{picture}(300,100)(0,0) \BCirc(50,50){35}
\BCirc(250,60){35} 
\Vertex(40,30){2} \Vertex(55,70){2} \Vertex(50,60){2}
\Vertex(65,50){2}
\Vertex(30,40){2}  \Vertex(30,62){2}
\Vertex(260,40){2} 
\Vertex(255,70){2} 
\Vertex(272,50){2}
\Vertex(232,40){2} 
\Vertex(232,60){2}
\Vertex(240,30){2}  
\Vertex(262,60){2} 
\ArrowLine(60,70)(150,70)
\ArrowLine(250,70)(150,70)
\Vertex(150,70){2}\ArrowLine(150,70)(100,50)\ArrowLine(150,70)(200,90)
\ArrowLine(45,30)(150,30) \ArrowLine(235,30)(150,30)
\Vertex(150,30){2}\ArrowLine(150,30)(100,20)\ArrowLine(150,30)(200,40)
\LongArrow(30,5)(100,5) \LongArrow(270,5)(200,5)
\end{picture}
\end{center}
\caption{Schematic view of double dijet prodution in a proton-proton
collision. }
\label{fig1}
\end{figure}

In high-energy hadronic collisions, more than one pair of partons can interact
with large momentum transfer. Such multiple hard interactions within the same
hadronic collision become more numerous with increasing center of mass energy. 
They are a novel and generic feature of hadronic interactions at Tevatron and at the LHC.
For instance, the inclusive cross section for double dijet production (see Fig.~\ref{fig1}) is 
$\sigma_D\left(E_{\rm T\, ,min} = 20\,  {\rm GeV}\right) \simeq 10\, \mu {\rm b}$ 
in proton-proton collisions at the LHC, if each of the four jets carries more than a minimal transverse
energy of $E_{\rm T\, ,min} = 20$ GeV. Even if this threshold 
is raised to  $E_{\rm T\, ,min} = 100$ GeV, the double hard scattering
process is still in experimental reach with 
$\sigma_D^{\rm 4\, jets}\left(E_{\rm T\, ,min} = 100\, {\rm GeV}\right) \simeq 50\, {\rm pb}$.
We will present  calculations, supporting these estimates, in section~\ref{sec2}.

It is not known how factorization theorems for large momentum transfer processes 
could be extended to multiple hard processes within the same hadronic
collision, such as double dijet production. Typically, one assumes that such processes 
can be described as the incoherent superposition of single hard scattering 
processes~\cite{Goebel:1979mi,Paver:1982yp,Halzen:1986ue,Mangano:1988sq,Godbole:1989ti}.
The double dijet cross section can then be expressed as the convolution of 
two independent hard partonic subprocesses with two-parton distribution functions $F_D$
(see eq. (\ref{eq2.4}) below). 
The ratio of the square of the dijet cross section $\sigma_S$  to the double dijet cross section $\sigma_D$ for two indistinguishable hard processes  defines the effective cross section $\sigma_{\rm eff}$~\cite{Paver:1982yp,Abe:1993rv,Abe:1997bp}
\begin{equation}
	\sigma_{\rm eff} = \frac{\sigma_S^2}{2\, \sigma_{ D}}\, .
	\label{eq1.1}
\end{equation}
For two distinguishable hard processes $A$ and $B$, it takes the form 
$\sigma_{\rm eff} = \frac{\sigma_A\, \sigma_B}{\sigma_{ D}}$. 
If the two-parton distribution functions $F_D$ factorize
into an uncorrelated product of  standard single parton distribution functions, then 
$\sigma_{\rm eff}$ gives access to the geometrical extension of the parton distributions 
in transverse space.  

There are several motivations for studying double hard cross sections at hadron colliders. 
First, hadronic collisions with more than one hard partonic scattering can contribute to
multi-parton final states at high transverse momentum. Their improved understanding 
may help to control the QCD background to searches for novel physics in
channels involving multiple high-$E_T$ parton final 
states~\cite{DelFabbro:1999tf}, although simple kinematic cuts can be
efficient in cleaning the signal  from double dijet
background~\cite{Godbole:1989ti}.
Second, multiple independent hard scatterings at lower momentum transfers
[$Q^2 \sim O\left( (1-5)^2\, {\rm GeV}^2\right)$] play an important role in modeling the
underlying event in hadronic collisions at collider 
energies~\cite{Sjostrand:1987su,Sjostrand:2004pf,Bahr:2008dy}. Studying the physics of 
such multiple hard scatterings at larger momentum transfer $Q^2$ or 
as a function of $Q^2$ may help to constrain
the input to this modeling of the underlying event~\cite{Drees:1996rw}.  Moreover,
the double dijet cross section $\sigma_D$ provides qualitatively novel information about 
the transverse structure 
of the hadronic projectile because $\sigma_D$ depends on the relative transverse 
distribution of the two partons in the hadronic projectile~\cite{Calucci:1999yz}. 
If the two partons were distributed
homogeneously over the entire hadronically active transverse area of the hadronic projectile, 
then the scale factor $\sigma_{\rm eff}$ 
would be comparable to the total inelastic cross section.
The much smaller numerical value  $\sigma_{\rm eff}=14.5\pm 1.7 ^{+1.7}_{-2.3}$ mb, 
measured by the CDF Collaboration~\cite{Abe:1997bp} disfavors such a 
homogeneous distribution. It is in support of a picture of
the proton and anti-proton, in which partons with large momentum fraction are localized
in a significantly smaller transverse region within the proton 
(see discussion of Fig.~\ref{fig2} below).

Quantum chromodynamics offers a specific picture for the transverse 
growth of hadronic wave functions with increasing $\ln 1/x$ or center of mass energy. 
In analogy to QED, where the transverse extension of the Weizs\"acker-Williams field
of quasi-real photons around an electric charge grows with increasing energy, the
hard (i.e. large-$x$) color charges in a QCD projectile can be viewed as
sources of non-abelian  Weizs\"acker-Williams fields, whose transverse size
grows with increasing $\ln 1/x$ or center of mass energy. The simplest, perturbative
realization of this phenomenon in QCD is the BFKL evolution equation, which predicts 
with increasing $\ln 1/x$ not only a growth of parton density locally in impact parameter,
but also a growth of the hadronic projectile distribution function in impact parameter
space, see e.g. 
Refs.~\cite{Gribov:1984tu,Salam:1995uy,Kovner:2002xa,Weigert:2005us,Shoshi:2002in}.
Unlike QED, this perturbative picture is expected to be modified in QCD 
by saturation effects, which tame the growth of parton density locally in impact
parameter, and by non-perturbative effects, which amputate the gluonic Weizs\"acker-Williams 
fields at a transverse distance set by confinement. One expects that this
reduces eventually the growth of the average transverse extension of 
non-abelian Weizs\"acker-Williams fields from the perturbatively predicted power-law
dependence $\propto x^{-\omega}$  to a non-perturbative logarithmic increase.
The scale and dynamics of the transition between perturbative and non-perturbative
regime, as well as the physics in the non-perturbative regime remain under debate.
All arguments indicate, however, that the transverse extension of the hadronic densities
continues to grow at ultra-relativistic center of mass energies, albeit possibly much weaker
than predicted perturbatively \cite{Gribov:1984tu,Weigert:2005us,Shoshi:2002in}. 

Here, we investigate to what extent this qualitative picture of the growth of transverse hadronic 
distributions could be tested by measuring double dijet production at Tevatron and at 
the LHC in the range of semi-hard 
momentum fractions, say $0.001 \lsim x \lsim 0.1$, and hard momentum transfers. 
The starting point of our work is the observation that 
measurements of the inclusive double dijet cross section at the LHC can be performed over
a wide range in $E_{\rm T,min}$. In the following, we quantify this range and we investigate
to what extent it provides access to a possible dependence of the scale factor
on the cm-energy and transverse momentum cut-off. On the basis
of these calculations, we assess the sensitivity of measurements of $\sigma_D$ to the
transverse growth of the distribution of semi-hard partons in the hadronic projectile. 


\section{The formalism}
\label{sec2}

We first discuss the formalism, on which the following calculations of double dijet production 
cross sections
are based. To leading order (LO), the single scattering cross section to produce two massless partons
of transverse energy larger than $E_{\rm T, min}$ reads
\begin{equation}
\sigma_S(E_{\rm T, min})=\int_{x_{\rm min}}^1 dx_1 \int_{
x_{\rm  min}/x_1}^1 dx_2 \int_{\hat t_-}^{\hat t_+}
d\hat t \sum_{ij} f_i(x_1,Q^2) f_j(x_2,Q^2)
\frac{d\sigma^{ij}}{d\hat t}\, ,
\label{eq2.1}
\end{equation}
where
\begin{equation}
\hat t_{\pm}=-\frac{\hat s}{2}\left(1 \pm
\sqrt{1-\frac{4\, E^2_{\rm T, min}}{\hat s}}\right),\quad\quad x_1\, x_2 \geq x_{\rm min}= \frac{4\, E_{\rm T,min}^2}{s}\quad \mbox{and}\quad Q^2=p_\perp^2.
\label{eq2.2} \label{eq2.14}
\end{equation}
The sum is over all possible quark and gluon 2$\to$2 scattering channels
$ij$. For the purpose of the present study, expression (\ref{eq2.1}) is a sufficiently
good approximation for the cross section of a dijet, produced in a single hard scattering event. 
We shall convolute LO parton-parton scattering cross sections $\frac{d\sigma^{ij}}{d\hat t}$ with 
the CTEQ6L set of parton distribution functions \cite{Pumplin:2002vw} for $f_i$, which have been optimized
for the use in LO calculations. The scale $Q^2$ in the PDFs and in $\alpha_s$ is set to the 
transverse momentum $p_\perp^2=ut/s$ for massless partons in the 2$\to$2 process.
We have checked that the results for $\sigma_S(E_{\rm T,min})$ obtained with this input
from Eq.(\ref{eq2.1})
are consistent with Pythia 6.419 \cite{Sjostrand:2006za}.

We calculate the inclusive dijet production cross section as an incoherent superposition 
of two hard scattering processes within
the same hadronic collision
\begin{eqnarray}
\label{sigmad}
	\sigma_D\left(E_{\rm T, min}\right) &=& \frac{1}{2}
	\sum_{ijkl}\int d{\bf s}_1\, d{\bf s}_2\, d{\bf b}\, \int dx_1dx_2dx_3dx_4 d \hat t_1d \hat t_2\, 
	F_D^{ik}(x_1,x_2;{\bf b}-{\bf s}_1,{\bf b}-{\bf s}_2)\, 
	\nonumber \\
	&& \qquad \qquad \times       F_D^{jl}(x_3,x_4;{\bf s}_1,{\bf s}_2)\, 
	\frac{d\sigma^{ij}}{d\hat t_1}\frac{d\sigma^{kl}}{d\hat t_2}\, .
\label{eq2.4}
\end{eqnarray}
Here, the sum $\sum_{ijkl}$ is over all parton species, which contribute to the two
partonic processes $i+j \to 2$ jets and $k+l \to 2$ jets.  The symmetry
factor $\frac{1}{2}$ accounts for the fact that both partonic processes are 
indistinguishable in the sense that there is no operational prescription which
establishes a one-to-one mapping between one dijet and one specific hard
partonic cross section in (\ref{eq2.4}). The Mandelstam variables $\hat t_1$ and $\hat t_2$
are defined for two partonic $2\to 2$ processes with incoming parton momentum 
fractions $x_1$, $x_3$  and $x_2$, $x_4$, respectively. 
We choose the kinematic boundaries in the integrals over the incoming parton momentum
fractions $x_1, ...\, , \, x_4$ 
 such that all outgoing partons carry more than a minimal transverse
energy $E_{\rm T, min}$. 
(In principle, more sophisticated kinematic boundaries could be implemented
e.g. to require different $E_{\rm T,min}$-values for 
both pairs of jets, but we shall not explore such
possibilities in the following.) 
The integral in equation (\ref{eq2.4}) includes the two transverse positions
${\bf s}_1$, ${\bf s}_2$, at which the two hard processes take place, and  the
impact parameter ${\bf b}$ of the hadronic collision. The spatial information about the partons is specified 
in the two-parton distribution functions
\begin{equation}
F_D^{ik}(x_1,x_2;{\bf b}_1,{\bf b}_2;Q_1^2,Q_2^2)\, ,
\label{eq2.3}
\end{equation}
which depend not only on the transverse momentum fractions $x_1$, $x_2$ and the
virtualities $Q_1^2$, $Q_2^2$ of both partons inside the hadron, but also on their 
transverse positions ${\bf b}_1$, ${\bf b}_2$.
In the following, we shall often use a simplified notation, in which the virtualities
are not written explicitly as arguments of $F_D$. For these
virtualities, we will always choose the squared transverse momenta in the corresponding
partonic $2\to2$ subprocess, as in (\ref{eq2.1}), (\ref{eq2.2}). 

\subsection{A factorized ansatz for two-parton distribution functions}
The discussion in this section is based on a class of models, which satisfy the 
factorized ansatz
\begin{equation}
	F_D^{ik}(x_1,x_2;{\bf b}_1,{\bf b}_2; Q_1^2,Q_2^2)
		= F^i(x_1,{\bf b}_1,Q_1^2)\, F^k(x_2,{\bf b}_2,Q_2^2)\, ,
		\label{eq2.5}
\end{equation}
where
\begin{equation}
	F^i(x,{\bf b},Q^2) = n(x,{\bf b})\, f^i(x,Q^2)\, .
	\label{eq2.6}
\end{equation}
A  set of correlated two-parton distributions, which do not satisfy (\ref{eq2.5}) will be
discussed in section~\ref{sec4}. In equation (\ref{eq2.6}), 
$n(x,{\bf b})$ denotes the density of partons in the transverse plane. It is
normalized to unity, 
$\int d{\bf b}\, n(x,{\bf b}) = 1$, so that $f^i(x,Q^2) = \int d{\bf b}\, F^i(x,{\bf b},Q^2)$
are the standard single parton distribution functions. For the class of models studied here,
the transverse part of the density does not depend on the parton species $i$.
The ansatz of Eq.~(\ref{eq2.6}) contains information about the 
average transverse distance of the partons from the center of the proton in transverse
space. If the partons are uncorrelated in impact parameter,
the average distance between the two partons satisfies 
$\langle \left( {\bf b}_1 - {\bf b}_2 \right)^2\rangle = 
\langle {\bf b}_1^2\rangle + \langle {\bf b}_2^2\rangle$. 
In this sense, the ansatz (\ref{eq2.5}) also contains information about the average 
transverse distance between the two partons. In addition, it allows for a 
non--trivial $x$-dependence
of the transverse size of the hadronic projectiles. 

Let us consider first the simple case of an $x$-independent density $n(x,{\bf b}) = n({\bf b})$.
In this case, the geometrical
information entering cross sections can be expressed in terms of the nucleon
overlap function
\begin{equation}
	T_{\rm NN}({\bf b}) = \int d{\bf s}\, n({\bf s})\, n({\bf b}-{\bf s})\, .
	\label{eq2.7}
\end{equation}
The normalization of $n$ implies that $\int d{\bf b}\, T_{\rm NN}({\bf b})=1$. For the
factorized ansatz (\ref{eq2.5}), the double dijet cross section then takes
the form
   \begin{equation}
	\sigma_D\left( E_{\rm T,min}\right) = 
	\left[ \sigma_S\left( E_{\rm T,min}\right)\right]^2 \,\frac{1}{2}
	\int d{\bf b}\, T^2_{\rm NN}({\bf b})\, ,
	\label{eq2.8}
\end{equation}
and the scale factor (\ref{eq1.1}) reads
\begin{equation}
	\sigma_{\rm eff} = \frac{1}{\int d{\bf b}\, T^2_{\rm NN}({\bf b})}\, .
	\label{eq2.9}
\end{equation}
In the general $x$-dependent case, the nuclear overlap function (\ref{eq2.7}) will depend on the
momentum fractions $x_1, ...\, , \, x_4$ 
of the partons in both hadrons. As a consequence, the scale
factor becomes a function of the center of mass energy $\sqrt{s}$ and the jet energy threshold
$E_{\rm T,min}$.

\subsection{Interpretation of the scale factor in the model (\ref{eq2.5}) for 
two-parton distributions}
\label{sec2b}
%
\begin{figure}[!h]
\includegraphics[width=0.6\textwidth]{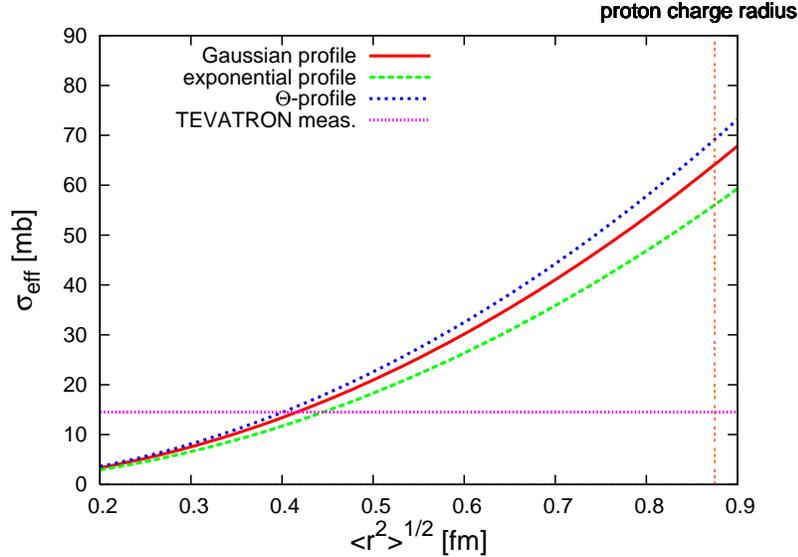}
\caption{ The scale factor $\sigma_{\rm eff}$ as a function of the 3-dimensional 
rms of the parton
density distribution for a Gaussian, exponential and $\Theta$-profile $n({\bf r})$.
The thinly dotted line denotes the central value of the Tevatron measurement
$\sigma_{\rm eff}=14.5$ mb. The value $\langle {\bf r}^2\rangle^{1/2} = 0.875\, {\rm fm}$
of the proton charge radius is also indicated. 
}
\label{fig2}
\end{figure}
%
For the class of models (\ref{eq2.5}), the scale factor contains
information about the
transverse parton density $n({\bf b})$ in the proton. A
model-independent understanding of the ${\bf b}$-dependence of this
density distribution is missing so far. Here, we consider
three-dimensional density profiles, from which transverse densities are
obtained by projection, $n({\bf b})=\int {\rm d}z\, n({\bf r})$. Rather
than motivating a particular ${\bf b}$-dependence of $n({\bf b})$,
however, we prefer here to demonstrate that our conclusions about the
scale factor will depend mainly on the rms of $n({\bf r})$ and will be
rather insensitive to details of the functional shape. To establish this
point, we compare three 3-dimensional parton densities $n({\bf r})$. In
particular, we consider a Gaussian density
\begin{equation}\label{eq2.10}
n({\bf r})=\frac{1}{(2\pi \delta^2)^{3/2}}\,\exp\left[-\frac{{\bf
r}^2}{2\delta^2}\right],
\end{equation}
a homogeneous density distribution which is sharply cut at radius $R$,
\begin{equation}\label{eq2.11}
n({\bf r})=\frac{3}{4\pi R^3}\,\Theta(R-|{\bf r}|)\, ,
\end{equation}
and an exponential profile
\begin{equation}\label{eq2.12}
n({\bf r})=\frac{1}{8\pi \lambda^3}\,\exp\left[-\frac{|{\bf
r}|}{\lambda}\right],
\end{equation}
which has a more pronounced tail than the Gaussian distribution.
To compare the sensitivity
of $\sigma_{\rm eff}$ on the functional shape of the ${\bf
r}$-dependence of the density profiles ($\ref{eq2.10}$),
($\ref{eq2.11}$) and ($\ref{eq2.12}$), we express the results in
Figure~\ref{fig2} and Table~\ref{tab1} in terms of the root mean square
$\langle {\bf r}^2\rangle^{1/2}$ of these densities.

One sees in Figure~\ref{fig2} that the numerical value of $\sigma_{\rm
eff}$ characterizes mainly $\langle {\bf r}^2\rangle^{1/2}$ and that its
sensitivity to the detailed geometrical profile of $n({\bf r})$ is
rather weak. As a consequence, the use of the Gaussian ansatz in the
following studies can be regarded as a convenient choice which will not
bias our conclusions. We mention as an aside that a Gaussian profile may
be motivated for instance by a study based on a light cone Hamiltonian
\cite{Pirner:2009zz}.
\renewcommand\arraystretch{2}
\begin{center}
\begin{table}
\begin{tabular}{|p{5cm}|p{3cm}|p{1.8cm}|p{2.5cm}|}\hline
Model for 3D density \hfill\hfill &\hfill ms radius $\langle {\bf r}^2 \rangle$ \hfill& \hfill$\sigma_{\rm eff}\hfill$ & \hfill$\langle {\bf r}^2\rangle_{\rm TEVATRON}\hfill$   \\ \hline
Gaussian $\propto\exp(-\frac{{\bf r}^2}{2\delta^2})$ & \hfill$3 \delta^2\hfill$ &  \hfill$\displaystyle\frac{8\pi}{3}\langle {\bf r}^2 \rangle\hfill$ &\hfill 0.42 fm \hfill\hfill \\
hard sphere  $\propto\Theta(R-|{\bf r}|)$& \hfill$\displaystyle\frac35 R^2\hfill$ & \hfill 9.0 $\langle {\bf r}^2 \rangle\hfill$ &\hfill 0.40 fm \hfill\hfill\\
exponential  $\propto\exp(-\frac{|{\bf r}|}{\lambda})$& \hfill 12$\lambda^2\hfill$& \hfill 7.3 $\langle {\bf r}^2 \rangle\hfill $ &\hfill 0.45 fm \hfill\hfill\\ \hline
\end{tabular}
\caption{\label{tab1} Results from the spatial density analysis. Calculations of ms radius $\langle {\bf r}^2 \rangle$ and $\sigma_{\rm eff}$ are shown for several density models. The last column is calculated by equating the $\sigma_{\rm eff}$ in the 4th column to the TEVATRON measurement of 14.5~mb. See also Fig.~\ref{fig2}.} 
\end{table}
\end{center}
Figure~\ref{fig2} also demonstrates that the central value $\sigma_{\rm eff} = 14.5\, {\rm mb}$ 
of the CDF measurement can be related to a narrow range around $\langle {\bf r}^2
\rangle^{1/2} \simeq 0.4-0.45\, {\rm fm}$ for all three geometrical profiles. 
We note that if the partons relevant for double dijet production were distributed in impact 
parameter over a transverse region of the size of the proton's electric 
charge $\langle {\bf r}^2\rangle^{1/2} = 0.875\, {\rm fm}$, then the scale factor 
$\sigma_{\rm eff}$ would take values between $50\, {\rm mb}$ and $70\, {\rm mb}$,
which are comparable to the total inelastic cross section $\sigma_{\rm inel} \simeq 80\, {\rm mb}$
measured at Tevatron.  A significant difference between the total hadronically active
transverse size of the proton and the transverse extension of the region relevant
for processes of high momentum transfer has been discussed repeatedly in
the literature~\cite{Strikman:2004km,Shoshi:2002in,Kopeliovich:2007pq}. 
We note that the measurement of $\sigma_{\rm eff}$ can not only provide an
independent characterization of this difference. Moreover,
it may also provide novel access to the dynamical origin of this difference
via an analysis of the
$x$-dependence of $\sigma_{\rm eff}$, to which we turn now.

\subsection{Modeling the small-$x$ evolution of the transverse size of hadronic wave functions}
\label{sec2c}

The transverse size of hard partonic components of the proton is found to be smaller than total cross sections but will also grow with increasing $\sqrt{s}$.
Therefore a larger part of the hadronically active regions in transverse space can be expected to 
contribute to hard production processes with increasing energy.

We  model this picture of the transverse growth of hadronic wave functions
by specifying an $x$-dependence of the Gaussian density (\ref{eq2.11}), 
\begin{equation}
	n(x,{\bf b})=\frac{1}{2\pi\delta(x)^2}
	\exp\left[-\frac{{\bf b}^2}{2\delta(x)^2}\right] \, .
		\label{eq2.11b}
\end{equation} 
We consider two parametrizations of the $x$-dependence of the Gaussian width $\delta(x)$.
One model, originally proposed by Burkardt \cite{Burkardt:2002hr}, takes
\begin{equation}
	\delta(x) = w_1\, \sqrt{(1-x)\, \ln\left(1/x\right)},\, \qquad\quad w_1=0.149\, {\rm fm}.
	\label{eq2.15}
\end{equation}
This results in a growth of the scale factor $\sigma_{\rm eff}(x) \propto \delta^2(x)
\propto \ln 1/x \propto \ln s$, which is even weaker than the growth 
$\propto \left( \ln s \right)^2$ which is realized in the Froissart bound. We 
fix the prefactor at $w_1=0.149\, {\rm fm}$ 
to reproduce  the Tevatron value for $\sigma_{\rm eff}$ at $E_{\rm T,min} = 20$ 
GeV.

We also consider a second model which results in a power-law growth of 
$\sigma_{\rm eff} \propto s^\omega$ with the center of mass energy,
\begin{equation}
	\delta(x) = w_2\, \left(1-x \right)\, x^{-\omega/2 },\, \qquad\quad w_2=0.175\, {\rm fm}\, .
	\label{eq2.16}
\end{equation}
A power-law growth of the transverse hadronic wave function is obtained in the
perturbative small-$x$ evolution, where the leading order BFKL-intercept is 
$\omega = \frac{\alpha_s}{\pi}\, N_c\, 4\, \ln 2$. This is known to overestimate the
growth of total cross sections in the experimentally
accessible regime. However, it is conceivable that the growth of hard components
in the transverse wave function is more rapid. Within the window of physically
reasonable parameters, we have chosen
$\omega = 0.265$ to arrive at a model with power-like growth at small $x$. The
prefactor in (\ref{eq2.16}) is fixed to $w_2=0.175\, {\rm fm}$ such that 
$\sigma_{\rm eff} = 14.5 \, {\rm mb}$ at $\sqrt{s}=1.8\, {\rm TeV}$ and
$E_{\rm T,min}=20\, {\rm GeV}$.

\begin{figure}
\includegraphics[width=0.6\textwidth]{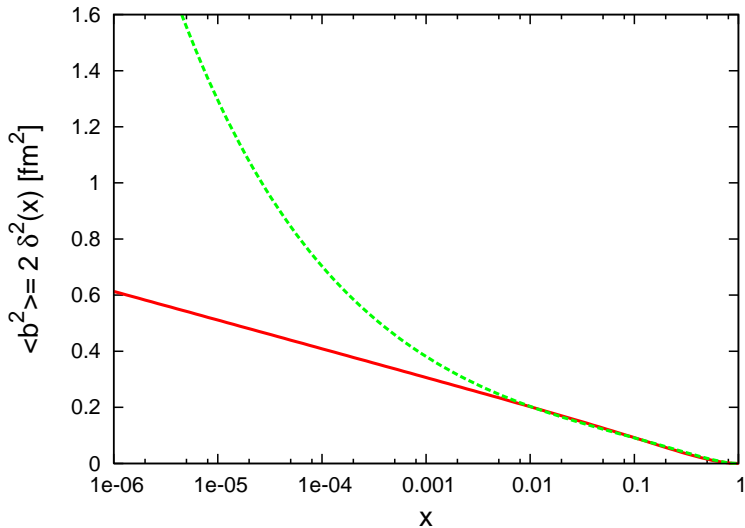}
\caption{Two models for the increase of $\langle {\bf b}^2\rangle=2\delta^2(x)$ 
towards small $x$ of the transverse proton wave 
function as a function of the parton momentum $x$. The lower (logarithmic) curve is
a model (\ref{eq2.15}) proposed by Burkardt \cite{Burkardt:2002hr}, the upper (power-like) curve 
is discussed in eq. (\ref{eq2.16}). 
}
\label{fig3}
\end{figure}

In Fig.~\ref{fig3} we show the transverse growth of the width
$\langle {\bf b}^2 \rangle=2 \delta^2(x)$ as a function of momentum fraction $x$ for the models (\ref{eq2.15}) and (\ref{eq2.16}). 
We note that a transverse growth similar to the model (\ref{eq2.15}) has also
been obtained in other calculations which model non-perturbative effects~\cite{Shoshi:2002in}.

A comment about our use of CDF measurements is needed here: 
in the CDF publication~\cite{Abe:1993rv} of 1993, a central value 
$\sigma_{\rm eff} = 12.1\, {\rm mb}$ was quoted for $E_{\rm T,min} = 18\, {\rm GeV}$ on
the parton level. The later CDF measurement~\cite{Abe:1997bp} quotes a central 
value $\sigma_{\rm eff} = 14.5\, {\rm mb}$
with much improved statistical and systematic uncertainties. However, both CDF measurements
considered a channel with three jets and one photon in the final state, while we focus on
four jet processes in the present discussion. The possibility that the scale factor differs for different
channels due to their dependences on different parton distribution functions, cannot be excluded
and has been explored in a recent model study~\cite{DelFabbro:2000ds}. To arrive at
a particularly simple model, and since there is little phenomenological guidance, we 
do not explore the possibility of such differences here. We also remark that
the more recent CDF analysis uses several lower values of $E_{\rm T,min}$. 
For these reasons, we emphasize that our choice
$\sigma_{\rm eff}\left(\sqrt{s}=1.8\, {\rm TeV}, E_{\rm T,min} = 20\, {\rm GeV}\right) = 14.5\, {\rm mb}$
does not reproduce the cuts and conditions of the 
CDF analysis. In particular, our choice of $\sigma_{\rm eff} = 14.5\, {\rm mb}$ can only be
related to the CDF analysis under the assumption that the scale factor does not depend on 
the production channel. Moreover, our choice of $E_{\rm T,min}$ is dictated by the need of anchoring our discussion at a sufficiently large transverse energy $E_{\rm T,min}$, where the perturbative cross sections used in our calculations are sufficiently reliable.

\section{Numerical results for $\sigma_{\rm eff}$ from double hard cross sections}
\label{sec3}
In this section, we first characterize the range of $E_{\rm T,min}$, which is 
experimentally accessible with sufficiently large event samples for the
study of double dijet production at Tevatron and at the LHC. We then discuss
the scale dependence of $\sigma_{\rm eff}$ in the model (\ref{eq2.15}) of
Burkardt and in the BFKL model (\ref{eq2.16}). 
Finally, we turn to the question to what extent more complicated geometrical
arrangements of two-parton distributions, or correlations not encoded for in the
ansatz (\ref{eq2.5}) can affect our conclusions. 
%

\subsection{Rate of inclusive double 2-jet processes}
\label{sec3a}

\begin{figure}[h!]
\includegraphics[width=0.495\textwidth]{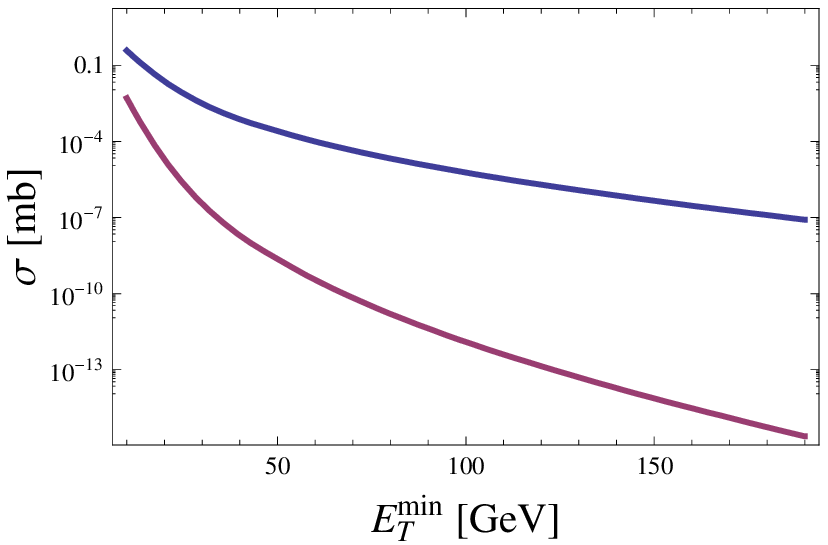}
\includegraphics[width=0.495\textwidth]{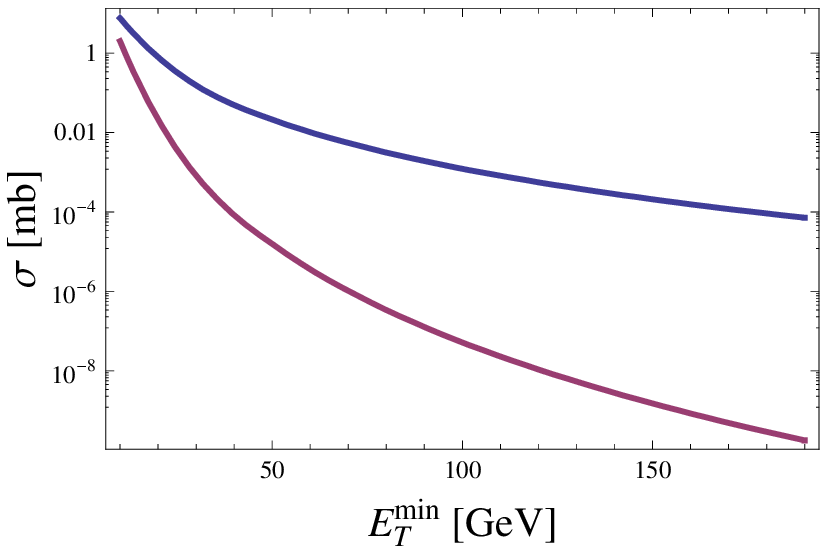}
\caption{Calculated inclusive single dijet cross section $\sigma_S$ (upper curves) and estimated double dijet cross
section $\sigma_{\rm D}=\sigma_{\rm S}^2/(2\, \sigma_{\rm eff})$ (lower curves) as a function of the cut on jet transverse energy.
Left hand side: $\sigma_{\rm S}$ and $\sigma_{\rm D}$ at $\sqrt{s}= 1.8$ TeV.
Right hand side: $\sigma_{\rm S}$ and $\sigma_{\rm D}$ at $\sqrt{s}=14$ TeV. 
The scale factor is assumed to be $\sigma_{\rm eff}=14.5$ mb, independent of $E_{\rm T,min}$ and $\sqrt{s}$. \label{fig4}
}
\raisebox{9.2cm}[4mm][2mm]{\hspace{2.5cm}Tevatron}
\raisebox{9.2cm}[4mm][2mm]{\hspace{7.5cm}LHC}
\raisebox{7.9cm}[4mm][2mm]{\hspace{-7cm}$\sigma_{\rm S}$} 
\raisebox{5.7cm}[4mm][2mm]{$\sigma_{\rm D}$}
\raisebox{7.9cm}[4mm][2mm]{\hspace{7cm}$\sigma_{\rm S}$}
\raisebox{5.9cm}[4mm][2mm]{\hspace{0cm}$\sigma_{\rm D}$}

\end{figure}

Figure~\ref{fig4} shows the calculated cross section $\sigma_S$ for inclusive 2-jet production 
and an estimate of the double dijet cross section $\sigma_D$ as a function of the minimal transverse 
energy of the jets.
To arrive at this estimate, the double dijet cross sections $\sigma_D$ in Figure~\ref{fig4}
is calculated from the single inclusive cross section $\sigma_S$ using equation (\ref{eq1.1})
with a scale-independent value $\sigma_{\rm eff} = 14.5\, {\rm mb}$. 
If $\sigma_{\rm eff}$ changes with
$\sqrt{s}$, then the inclusive double 2-jet cross section $\sigma_D$ would differ from
the value shown in Fig.~\ref{fig4} by a factor $14.5\, {\rm mb}/\sigma_{\rm eff}(\sqrt{s})$. 
A table with numerical values for double dijet cross sections, entering Figures~\ref{fig4}
and ~\ref{fig5}, is provided in the electronic supplement to this paper. 

At the Tevatron Run I ($\sqrt{s}= 1.8$ TeV), the 
inclusive double dijet 
cross section reaches $\approx 20\, {\rm nb}$ for $E_{\rm T,min} \simeq 20$ GeV.
Upon increasing the jet energy threshold, this cross section drops rapidly to
$\approx 20\, {\rm pb}$ for $E_{\rm T,min} \simeq 40$ GeV. The kinematical reach
at the LHC ($\sqrt{s}=14$ TeV) is much wider. With a cross section of 
$\approx 16\, {\rm nb}$, one gets to $E_{\rm T,min} \simeq 50$ GeV, and with a cross
section of $10\, {\rm pb}$, one explores the scale dependence of 
$\sigma_D$ up to  $E_{\rm T,min} \simeq 120$ GeV. 

To put these cross sections into perspective, let us assume an integrated luminosity
for Run II at Tevatron of $10\, {\rm fb}^{-1}$ (more than $6\, {\rm fb}^{-1}$ have been
delivered to date). This would translate into $2\times 10^8$ double dijet events with $E_{\rm T,min} \simeq 20$ GeV and $2\times 10^5$ events with 
$E_{\rm T,min} \simeq 40$ GeV.
(For the purpose of these order of magnitude estimates, we have neglected the difference
in center of mass energy between Tevatron Run I and Run II.)
To relate the double hard cross section $\sigma_D$ 
to a measurable quantity, it is necessary to disentangle 
the 4-jet events originating from two independent hard scattering processes from 
those that originate from a single hard scattering. 
There are various experimental handles for doing this. For instance, one can
exploit that  in contributions to $\sigma_D$,
pairs of two jets must be balanced in $E_T$, while contributions from other classes of
4-jet events are not and differ in the shape of their distribution. However,
our study does not provide a basis for judging the (experiment-specific) size of
event samples, needed for a measurement of $\sigma_D$. Here, we assume that $\sim 10^5$
raw events are sufficient to this end, and we thus estimate that experiments at
Tevatron Run II can measure $\sigma_D$ for $E_{\rm T,min} \lesssim 40$ GeV.

Under analogous assumptions, experiments at the LHC will access the physics of
double dijet production for  $E_{\rm T,min} \lesssim 120$ GeV
with the first $10\, {\rm fb}^{-1}$ of data at $\sqrt{s}=14\, {\rm TeV}$. 

\subsection{The scale dependence of the scale factor $\sigma_{\rm eff}$}
\label{sec3b}
\begin{figure}[h!]
\includegraphics[width=0.495\textwidth]{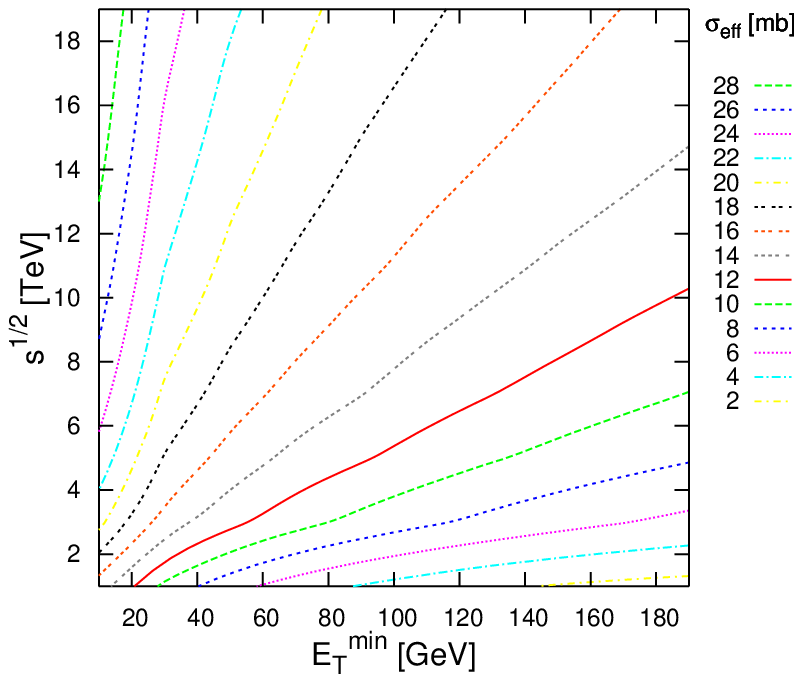}
\includegraphics[width=0.495\textwidth]{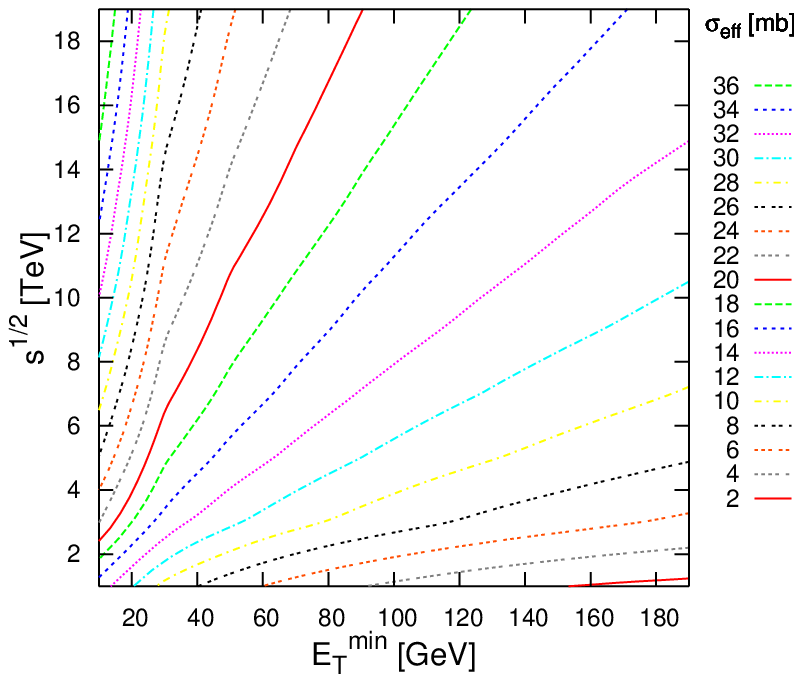}
\caption{Theoretical calculation of the  scale factor $\sigma_{\rm eff} = \sigma_S^2\left(E_{\rm T,min},\sqrt{s}\right)/
2\, \sigma_D\left(E_{\rm T,min},\sqrt{s}\right)$ in the plane of 
center of mass energy $\sqrt{s}$ and jet energy threshold $E_{\rm T,min}$. Constant $\sigma_{\rm eff}$-values indicated on the right of the two plots lie on approximately straight lines.
Inclusive single and double dijet cross sections are calculated from (\ref{eq2.1}) 
and (\ref{eq2.3}) respectively, for a Gaussian transverse density distribution of 
partons in the proton wave function. The $x$-dependence of the width of these 
density distributions is taken to follow a logarithmic increase (\ref{eq2.15}) [plot 
on left hand side] or a  power-law increase (\ref{eq2.16})
[plot on right and side]. Model parameters are chosen such that 
$\sigma_{\rm eff} = 14.5\,  {\rm mb}$ for $\sqrt{s} = 1.8\, {\rm TeV}$ and 
$E_{\rm T, min} = 20\, {\rm GeV}$.
}
\label{fig5}
\end{figure}

The estimates given in section~\ref{sec3a} above illustrate that experiments at the LHC
can investigate the scale dependence of the scale factor $\sigma_{\rm eff}$ over a
logarithmically wide range in $E_{\rm T,min}$. We now illustrate the physical information
which can be extracted from this scale dependence. To this end, 
we have calculated the inclusive double dijet cross section (\ref{eq2.3}) for two-parton
distribution functions (\ref{eq2.5}) with Gaussian transverse density profile $n(x,{\bf b})$. 
Figure~\ref{fig5} shows results for the corresponding scale factor $\sigma_{\rm eff}$,
calculated as a function of $\sqrt{s}$ and $E_{\rm T,min}$ for the cases of a
logarithmic and a power law $x$-dependence of the width $\delta(x)$ of the parton density
$n(x,{\bf b})$ in the hadronic projectile.

We expect that experimental data on $\sigma_{\rm eff}$ at the LHC will first become
available as a function of $E_{\rm T,min}$ for fixed $\sqrt{s}$. However, the foreseen
running schedule of LHC may also lead to information about the $\sqrt{s}$-dependence
of double dijet cross sections. This is so, since before moving to $\sqrt{s}=14\, {\rm TeV}$,
LHC is scheduled to start operation this year with $\sqrt{s}= 10$ TeV, aiming for an
integrated luminosity of $200\, {\rm pb}^{-1}$ which may be sufficient to explore
double hard collisions over a range in $E_{\rm T,min}$. Moreover, at a 
later stage in the LHC program, one may also expect a relatively short proton-proton run at 
$\sqrt{s}= 5.5$ TeV to collect comparison data for the LHC heavy ion program.
In addition, data from Tevatron Run II may provide information about the $E_{\rm T,min}$
dependence of $\sigma_D$ at $\sqrt{s}=1.96$ TeV.
Measurements of the $\sqrt{s}$-dependence of $\sigma_D(E_{\rm T,min})$ will test the
constancy of $\sigma_{\rm eff}$ along lines of constant $E_{\rm T,min}/\sqrt{s}$.
This constancy does not depend on details of the modeling of 
two-parton distribution functions, but results solely from the kinematic bound (\ref{eq2.14}).
This makes it an important consistency check for the picture of double dijet production
advocated here. 
%
\begin{figure}
\includegraphics[width=0.6\textwidth]{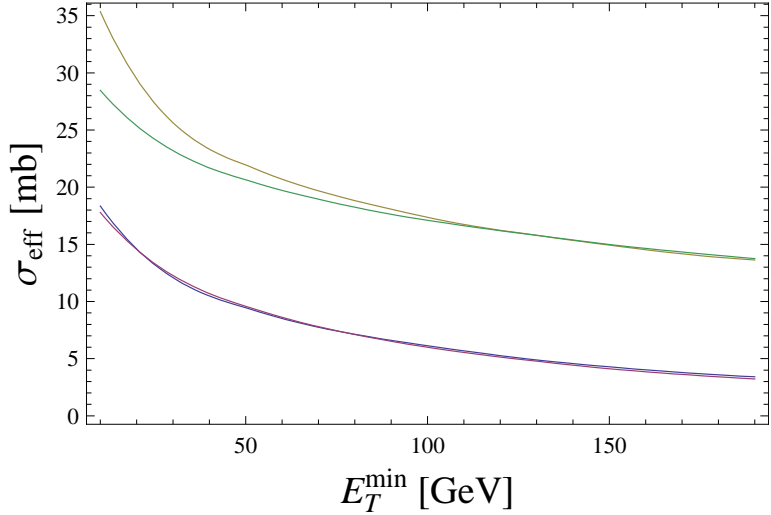}
\caption{The scale factor $\sigma_{\rm eff}$ for the two models of an
$x$-dependent width of the Gaussian density distribution, shown in Fig.~\ref{fig3}.
Upper curves for $\sqrt{s} = 14\, {\rm TeV}$, lower curves for $\sqrt{s}=1.8\, {\rm TeV}$.}
\label{fig6}
\end{figure}
%
The main dynamical information of the measurement of $\sigma_{\rm eff}$ is contained
in its dependence on $E_{\rm T,min}$ and $\sqrt{s}$. The variation of this scale factor
with kinematical variables reflects the growth of the transverse size of the projectile
wave function with $\ln 1/x$. Fig.~\ref{fig5} illustrates this point for the two models of
the transverse growth represented in Fig.~\ref{fig3}. One sees that the
scale factor increases significantly with increasing $\sqrt{s}$ or 
decreasing $E_{\rm T,min}$.
We emphasize that at $\sqrt{s}=14$ TeV, results of both models shown in 
Fig.~\ref{fig5} imply a variation of $\sigma_{\rm eff}$ by roughly a factor $2$ in the
range between $E_{\rm T,min}=10$ GeV and $E_{\rm T,min}=100$ GeV. This
variation is much larger than the $\sim 20\%$ measurement uncertainty quoted 
by the CDF Collaboration for its measurement of $\sigma_{\rm eff}$. We take this
as a strong indication that the $x$-evolution of the transverse size of hadronic 
wave functions in the range of semi-hard $x$ is experimentally accessible via
the measurement of double hard cross sections at the LHC.

Once a non-trivial $E_{\rm T,min}$-dependence of the scale factor $\sigma_{\rm eff}$
is established, one may ask the refined question of whether this allows for the
discrimination between different models of small-$x$ evolution. Since parton distributions $f(x,Q^2)$ rise rapidly with increasing $\ln 1/x$, the inclusive
single and double dijet cross sections $\sigma_S(E_{\rm T,min})$, $\sigma_D(E_{\rm T,min})$
are dominated by $x$-values which lie close to the lower bound of (\ref{eq2.14}).
As a consequence, the value of $\sigma_{\rm eff}$ remains almost constant along lines 
of constant $E_{\rm T,min}/\sqrt{s}$ in the $\left(E_{\rm T,min},\sqrt{s}\right)$ plane. 
As an estimate, one may take 
\begin{equation}
\sigma_{\rm eff} \simeq 15\, \delta^2(x_{\rm min}^{1/2}),
\end{equation}
where $x_{\rm min}$ is the lower bound on the parton momentum fraction available
for a dijet above $E_{\rm T,min}$ (see Eq. \ref{eq2.14})). 
Figure~\ref{fig6}
shows in more detail that for the two models of $x$-evolution shown in Fig.~\ref{fig3},
the steeper $x$-dependence in the transverse density profile $n(x,{\bf b})$
is indeed reflected in a steeper $E_{\rm T,min}$-dependence of $\sigma_{\rm eff}$.
The model-dependent difference is more pronounced for the higher LHC center of 
mass energy and for lower values of $E_{\rm T,min}$, where double hard scattering 
processes depend on parton distributions at smaller momentum fraction $x$.
However, the differences are rather mild and may be difficult to disentangle experimentally.
Moreover, the interpretation of relatively small variations in $\sigma_{\rm eff}$ may
require a more detailed understanding of the geometrical distributions entering $F_D$
(see section~\ref{sec4} below).
Thus, while a non-trivial $E_{\rm T, min}$-dependence of $\sigma_{\rm eff}$ will
allow one to disentangle models of small-$x$ growth from the baseline of an
$x$-independent transverse localization of partons in the proton, the discrimination
between different models of small-$x$ growth may be more challenging.

\section{Correlated two-parton distributions}
\label{sec4}
So far, we have discussed factorized two-parton distributions of
the form (\ref{eq2.5}). This ansatz is based on a picture, in
which partons are centered around a single position ${\bf b}_v
\equiv 0$ in transverse space. The purpose of this section is to
gain some understanding of the extent to which the results reached
above depend on the geometrical assumptions underlying the ansatz
(\ref{eq2.5}), and to what extent they reflect dynamical
information. To this end, we shall study a simple model of two-parton distributions,
which do not factorize into single-parton distributions. The model is based
on picturing the transverse profile of the proton
projectiles as being composed of three regions of size $\sim
0.2-0.4\, {\rm fm}$ each, which have increased hadronic interaction probability.
The centers ${\bf b}_{v_i}\, , i =1,2,3$, of these hot spots may
be thought to be related to the positions of valence quarks in the
transverse plane.  To be specific, we consider for these hot spots 
the distribution~\cite{Bender:1983cw}
\begin{equation}
|\psi({\bf b}_{v_1},{\bf b}_{v_2}|^2
=\frac{3}{\pi^2\delta_v^4}
\exp\left[-\frac{1}{3\delta_v^2}\left(({\bf b}_{v_1}-{\bf
b}_{v_2})^2+({\bf b}_{v_1}-{\bf b}_{v_3})^2+({\bf b}_{v_2}-{\bf
b}_{v_3})^2\right)\right] 
\Bigg\vert_{ -{\bf b}_{v_3} \equiv {\bf b}_{v_1} + {\bf b}_{v_2}}
\label{eq4.1}
\end{equation}
Here due to the center of mass constraint the third coordinate is
defined as ${\bf b}_{v_3} \equiv -{\bf b}_{v_1} - {\bf b}_{v_2}$.
We fix $\delta_v = 0.25\, {\rm fm}$,
which corresponds to a separation of the centers of the
three hot spots by an rms of 0.3 fm. Partons
are located around the centers  ${\bf b}_{v_i}$
of these hot spots at transverse positions ${\bf b}$ with Gaussian
distributions
\begin{eqnarray}
d(x,{\bf b},{\bf b}_v)=\frac{1}{2\pi\delta_s(x)^2}
\exp\left(-\frac{({\bf b}_{v}-{\bf b})^2}{2\delta_s(x)^2}\right)\, .
\label{eq4.2}
\end{eqnarray}
We allow for an $x$-dependence of the Gaussian width $\delta_s$. 
A two-parton density distribution $n_D(x_1,x_2,{\bf
b}_1,{\bf b}_2)$ for partons located at transverse positions ${\bf
b}_1$, ${\bf b}_2$ can then be calculated in terms of the
following integral over the centers of the hot spots ${\bf
b}_{v_i}$,
\begin{eqnarray}
n_D({x_1,x_2;\bf b}_1,{\bf b}_2) &=& \frac{1}{4}\int d{\bf
b}_{v_1}\, d{\bf b}_{v_2} \vert\psi\left({\bf b}_{v_1},{\bf
b}_{v_2}\right)\vert^2\, \sum_{ij}^2 d(x_1,{\bf b}_1,{\bf
b}_{v_i})\, d(x_2,{\bf b}_2,{\bf b}_{v_j}). \label{eq4.3}
\end{eqnarray}
Here, the sum over $i$ and $j$ averages over the different
combinations of hot spots in one proton which can provide the two
partons for the hard scattering process. In this class of models,
the two-parton distribution functions do not factorize, but take
the form
\begin{equation}
  F_D^{ik}\left(x_1,x_2;{\bf b}_1,{\bf b}_2; Q_1^2,Q_2^2 \right)
      = n_D(x_1,x_2;{\bf b}_1,{\bf b}_2)\, f^i\left(x_1,Q_1^2\right)\,  f^k\left(x_2,Q_2^2\right)\, .
      \label{eq4.4}
\end{equation}
The corresponding integration over transverse directions, entering (\ref{eq2.4})
reads
\begin{eqnarray}
  &&\int d{\bf b}\, d{\bf s}_1\, d{\bf s}_2\,
      n_D\left(x_1,x_2; {\bf b}-{\bf s}_1,{\bf b}-{\bf s}_2\right)\,
          n_D\left(x_3,x_4; {\bf s}_1,{\bf s}_2\right)
          \nonumber \\
          && \qquad = \frac{1}{8\pi}\left( \frac{1}{\delta_\Sigma
^2}+\frac{1}{\delta _\Sigma ^2+2\delta_v^2}+\frac{2}{\delta
_\Sigma ^2+\delta_v^2} \right)\, , \label{eq4.5}
\end{eqnarray}
where
\begin{equation}
  \delta_\Sigma^2=\delta_s(x_1)^2+\delta_s(x_2)^2+\delta_s(x_3)^2+\delta_s(x_4)^2\, .
  \label{eq4.6}
\end{equation}
In the limit $\delta_v \to 0$, one recovers the model of a density
distribution with Gaussian profile, discussed in
section~\ref{sec3}. More precisely, for $\delta_v\to 0$, the
centers of the three hot spots are all located at ${\bf b}_{v_1} =
{\bf b}_{v_2} = {\bf b}_{v_3} = 0$, as can be seen from
(\ref{eq4.1}). For the case of an $x$-independent density
$\delta_s(x)=\delta_s$, Eq. (\ref{eq4.5}) becomes $\frac{1}{8\,
\pi} \frac{4}{4\, \delta_s^2} = \frac{3}{8\, \pi\, \langle {\bf
r}^2\rangle}$ with $\langle {\bf r}^2\rangle = 3\, \delta_s^2$.
This is exactly the value of $\int d{\bf b}\, T_{NN}^2({\bf b}) =
1/\sigma_{\rm eff}$ for the Gaussian profile, obtained in
Tab.~\ref{tab1}.

To study the effect of non-factorizing distributions and compare
to the results of section \ref{sec3} we make use of the Gaussian
width $\delta_{\rm eff}^2=\delta_s(x)^2+\delta_v^2/3$ of the
corresponding one-particle distribution $n(x_1,b_1)=\int d{\bf
b}_2\, n_D(x_1,x_2; {\bf b}_1,{\bf b}_2)$. We connect the
correlated one-particle distribution of soft partons to the
uncorrelated one by demanding that they have the same Gaussian
width, $\delta(x)=\delta_{\rm eff}$. This yields $\delta_s$ used
for the numerical computation.
\begin{figure}[h!]
\includegraphics[width=0.495\textwidth]{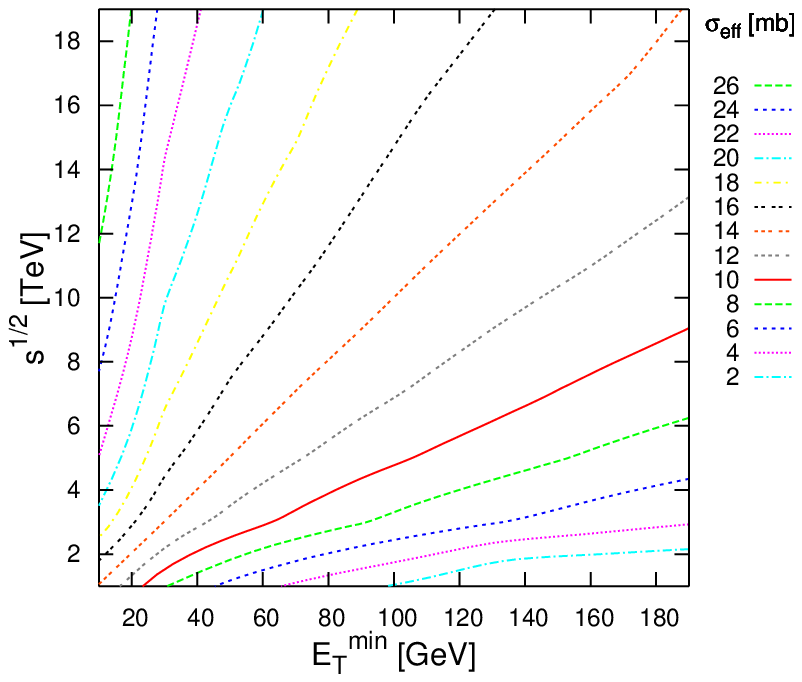}
\includegraphics[width=0.495\textwidth]{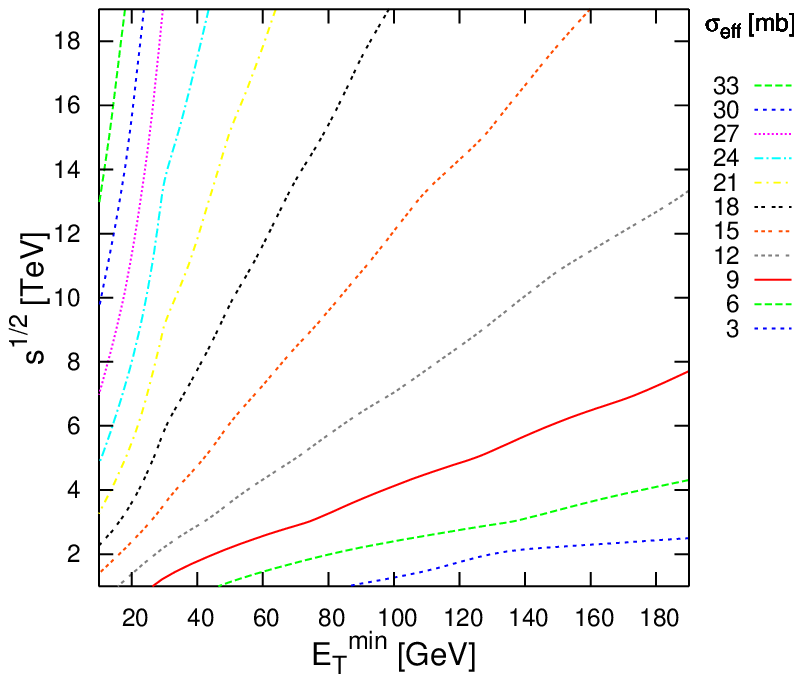}
\caption{Same as Fig.~\ref{fig5}, but for a model of two-parton distribution functions,
for which partons are localized in three hot spots in the parton wave function. }
\label{fig7}
\end{figure}

We have calculated the inclusive double dijet cross section (\ref{eq2.5}) for the correlated
two-parton distribution function (\ref{eq4.4}) and compare again a power-law $x$-dependence of the form (\ref{eq2.16}) with a logarithmic
$x$-dependence of the form (\ref{eq2.15}).
The results of this calculation are shown in Figure~\ref{fig7}.

Similar to the calculation with factorized two-parton distribution functions (\ref{eq2.5}) 
shown in Fig.~\ref{fig5}, we observe from Fig.~\ref{fig7} that the scale factor 
$\sigma_{\rm eff}$ increases with increasing $\sqrt{s}$ or decreasing $E_{\rm T,min}$. 
The numerical differences  between the results shown in Fig.~\ref{fig7} and
Fig.~\ref{fig5} are relatively small. This may be understood by observing that
the average distance between the centers of the hadronically active regions is
$\sqrt{\langle \left( {\bf b}_{v_1} - {\bf b}_{v_2} \right)^2 \rangle} = \sqrt{2}\, \delta_v
\approx 0.35\, {\rm fm}$ for $\delta_v = 0.25$ fm. The average distance from the
center ${\bf b}_{v_i}$, at which partons are localized is
$\sqrt{\langle {\bf b}^2 \rangle} = \sqrt{2\, \left(\delta_v^2/3 + \delta_s^2(x) \right)}
\geq \sqrt{2/3} \delta_v$. This is  $\approx 0.20$ fm at large $x$ and increases
significantly for smaller $x$. As a consequence, the three hadronically active
regions in the model (\ref{eq4.3}) 
overlap significantly for our choice of model parameters, and measurable 
properties of this distribution are likely to be similar to those of the single
homogeneous density distribution, studied in section~\ref{sec3}.

\section{Discussion}

In this paper, we have argued that the proton wave function is expected to grow in
the transverse plane with increasing $\ln 1/x$ for all values of $x$. We have then 
illustrated in model studies that this transverse growth should manifest itself in two-parton distribution functions and that it becomes experimentally accessible in the inclusive double 
dijet cross section $\sigma_D$.
Of particular interest is the ratio $\sigma_{\rm eff} = \sigma_S^2/2\, \sigma_D$ of the
square of the inclusive single over the double dijet cross section. As a generic 
consequence of the picture advocated here, the scale factor $\sigma_{\rm eff}$ is 
expected to be constant along lines of constant $E_{\rm T,min}/\sqrt{s}$, and it is 
expected to grow with $\sqrt{s}$ at fixed jet energy threshold $E_{\rm T,min}$ and to 
decrease with increasing  $E_{\rm T,min}$ at fixed $\sqrt{s}$. We have shown that
the wider kinematical reach of proton-proton collisions at the LHC will allow for the first time to
test this scale dependence of the scale factor $\sigma_{\rm eff}$ over a wide range in $E_{\rm T,min}$.

We have studied models for parton distributions in the transverse plane, which do not 
distinguish between
gluons and valence and sea quarks of different flavor. In the absence of experimental information
about such differences, we have adopted the baseline assumption that the geometrical 
distributions of all partons are the same, thus minimizing the number of model parameters.  
We note, however, that one may invoke QCD-inspired pictures for which different parton species
are localized differently in the transverse wave function. In this case, the scale factor 
$\sigma_{\rm eff}$
can depend on the production channel of the double hard scattering process, since different production
channels (such as final states with two-jet events or with $b\, \bar{b}$) depend on different 
parton densities. A model, which shows this feature, was studied for instance by Del Fabbro and
Treleani in Ref.~\cite{DelFabbro:2000ds}. The study of specific double hard production channels 
has also been explored  as a means to arrive at an improved (ideally: background free) 
experimental characterization of double parton collisions. In particular, the production of two
equal sign W bosons at relatively low transverse momentum is dominated by double parton
collisions~\cite{Kulesza:1999zh}, though the cross section is very low. A much
more abundant channel at the LHC, even if $b$-tagging efficiency is taken into account, is the production of
$b\, \bar{b}\, b\, \bar{b}$, which may allow for an improved characterization of double hard
cross sections~\cite{TomLeCompte}.

The models explored in the present paper result in an increase of $\sigma_{\rm eff}$ with
increasing $\sqrt{s}$. In contrast, the spatial correlations, implemented in the model of 
Ref.~\cite{DelFabbro:2000ds}, imply that scale factors for all production channels 
decrease with increasing $\sqrt{s}$. 
This illustrates  that the models studied in the present work do not exhaust 
all conceivable scale dependencies of $\sigma_{\rm eff}$. The models in
section~\ref{sec3} and ~\ref{sec4} are  simple implementations of the picture that the 
transverse proton wave function grows with $\ln 1/x$.
Moreover, the difference between our models and the models of Ref.~\cite{DelFabbro:2000ds} 
illustrates clearly, that far beyond producing only an experimental value for $\sigma_{\rm eff}$,
a measurement of inclusive double hard cross sections at the LHC can distinguish between
qualitatively different pictures of the transverse proton wave function. As we have argued
here, this may be done most efficiently by studying the $E_{\rm T,min}$-dependence of the
scale factor $\sigma_{\rm eff}$ at fixed $\sqrt{s}$. 

\acknowledgements
We thank Tom LeCompte, Tilman Plehn, Steffen Schumann and Daniele Treleani
for useful discussions. 
S.D. has been supported within the framework of the Excellence Initiative by the
German Research Foundation (DFG) through the Heidelberg Graduate School of
Fundamental Physics (grant number GSC 129/1).


\end{document}